\documentclass[11pt,twoside]{article}

\usepackage{asp2004}
\usepackage{epsf}
\usepackage{psfig}
\usepackage{lscape}

\begin{document}
\title{Optical Monitoring of PKS 2155-304 during August-September 2004 with the KVA telescope}   

\author{S. Ciprini,  E. Lindfors, K. Nilsson, L. Ostorero}   

\affil{Tuorla Observatory, University of Turku, V\"{a}is\"{a}l\"{a}ntie 20,\\ 21500 Piikki\"{o}, Finland}    

\begin{abstract} 
The southern gamma-ray blazar PKS 2155-304 is one of the brightest
and most intensively studied prototypes of BL Lac object. PKS
2155-304 has recently aroused the interest of Cherenkov telescope
projects like HESS and MAGIC, the former having already observed
the source in 2002 and 2003. This blazar was monitored with the
KVA optical telescope (R-band intranight photometry and unfiltered
polarization observastions), in the frame of a new HESS
multiwavelength campaign performed in August-September 2004.
\end{abstract}



\vspace{-5mm}%
\section{Introduction}
\begin{figure}[t!!]
\centering
\hspace{-0.5cm} %
\epsfysize=5.8cm  \epsfbox{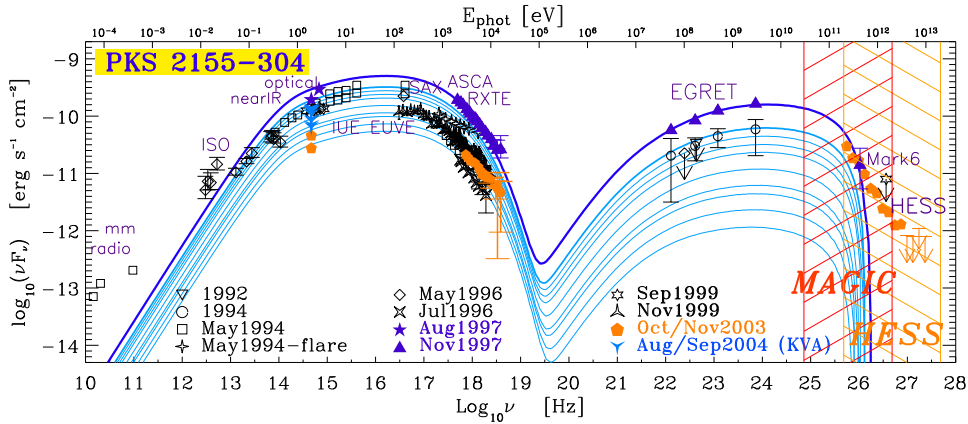}
\vspace{-5mm} \caption[h]{The SED of PKS 2155-304 assembled with
observations available around mid-90s \citep[the solid lines are a
SSC cooling modelling of Nov.1997 data,][]{ciprini03}, and with
data from the HESS campaign of Oct.-Nov.2003 \citep{Aharonian05b}
for comparison. The qualitative sensitivity energy ranges of MAGIC
and HESS are superimposed. EGRET energy spectrum has an integral
photon index of $1.71 \pm 0.24$, while the time-averaged photon
index measured by HESS in 2002-2003 is $3.32 \pm 0.06$
\citep{Aharonian05a}.}\label{figSED} \vspace{-6mm}
\end{figure}

The southern BL Lac object PKS 2155-304 (z = 0.117) is one of the
brightest extragalactic UV and X-ray sources in the sky. This
blazar exhibits a complex broadband variability and has been the
target of several multiwavelength campaigns and many X-ray
satellites like ROSAT, ASCA, RXTE, SAX, Chandra, and XMM
\citep[see
e.g.][]{vestrand99,chiappetti99,brinkmann00,kataoka00,edelson01,tanihata01,zhang02,
nicastro02,ciprini03,cagnoni04}. Fig.\ref{figSED} shows an example
of its spectral energy distribution (SED), peaking in the
UV-soft-X-ray regime. A continuous X-ray flaring activity,
representing the high-energy tail of the synchrotron emission, is
typical of this source \citep{tanihata01}. PKS 2155-304 was well
observed by EGRET in the $\gamma$-ray energy range 30 MeV -- 10
GeV \citep[see e.g.][]{urry97,vestrand99}, and was also detected
by the Cherenkov telescope Mark-6 (Australia) in 1997
\citep{chadwick99}. It was not revealed by the same instrument and
by the CANGAROO telescope in subsequent years; it was finally
detected with high-significance ($45\sigma$) by the HESS array of
Cherenkov telescopes (Namibia) in 2002--2003 \citep[][see
Fig.\ref{figSED}]{Aharonian05a,Aharonian05b} and in 2004
(Aharonian et al. in prep). Radio observations (sparse if compared
to the optical and high-energy ones) show no strong moving
components in VLBI, suggesting low Doppler factors and a mildly
relativistic jet on parsec-scale \citep{piner04}.
%
%
\section{Optical Observations with the KVA Telescope}
Observations of very high energy $\gamma$-rays from PKS 2155-304
by Cherenkov telescopes like HESS
(\texttt{http://www.mpi-hd.mpg.de/hfm/HESS/}) and MAGIC
(\texttt{http://wwwmagic.mppmu.mpg.de}) simultaneous with optical
observations, can provide a noticeable insight into TeV emission
processes and help to clarify the differences between blazars
peaked at low and high-energy. The recent (1996-2000) optical
history of PKS 2155-304 exhibits quiescent phases of mild
variability followed by periods of more rapid and stronger
activity \citep[day timescales; ][]{dominici04}, while intranight
photo-polarimetric data show a moderate polarization degree
\citep{tommasi01}. The simultaneous optical and X-ray (RXTE-ASM)
emission does not show any correlation \citep{dominici04}.
\par The KVA telescope (Kungliga Vetenskapsakademien, Royal
Swedish Academy of Sciences) is located on Roque de los Muchachos,
La Palma (Canary Islands), and operated by the Tuorla Observatory,
Finland (\texttt{http://www.astro.utu.fi}). KVA observations are
the main contribution of the Tuorla AGN group to the MAGIC
collaboration. MAGIC is a 17m atmospheric imaging Cherenkov
telescope (located in the same site), designed to detect cosmic
$\gamma$-rays with a low-energy threshold \citep[$\sim$40 GeV, see
e.g. ][]{lorenz04,cortina05}.
%
\begin{figure}[t!!]
%
\hspace{-0.3cm} \epsfysize=13.5cm
\rotatebox{90}{\epsfbox{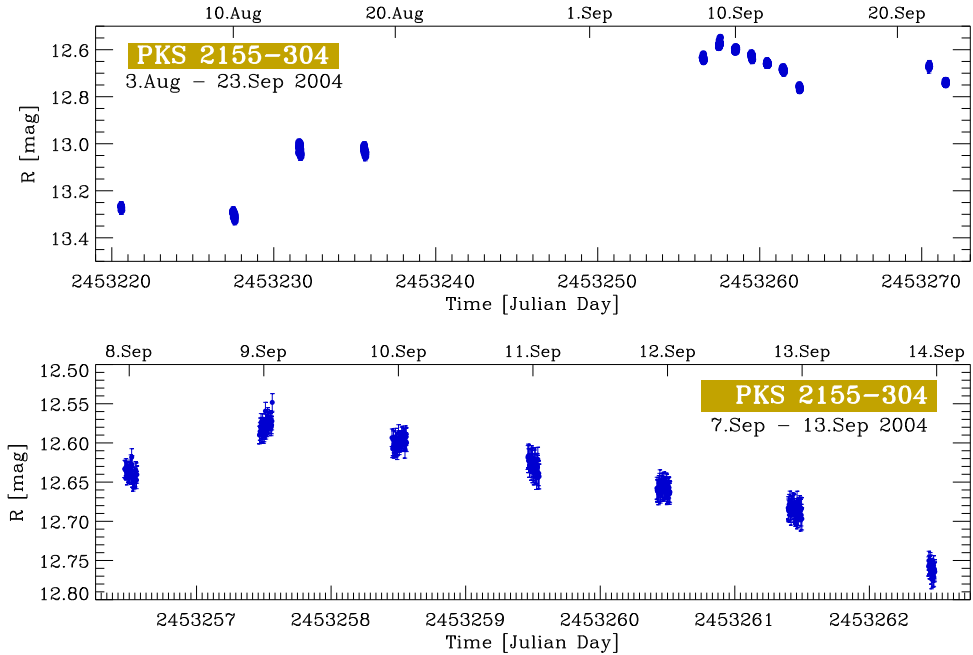}}
\vspace{-8mm} \caption[h]{Optical light curve of PKS 2155-304 in
$R$-band obtained with KVA observations from August 3 to September
23, 2004, during a contemporary multiwavelength campaign involving
HESS and RXTE. A total of 440 data points were obtained in 13
nights ($\sim 38$ frames per night on average). The peak of the
optical emission is located at JD=2453257.5627 (Sep. 9) with
$R=12.548 \pm 0.012$ mag (unabsorbed flux $31.08 \pm 0.32$ mJy).
The fairly well sampled week around this flare is plotted in the
bottom panel.}
\label{figlightcurves} \vspace{-5mm}
\end{figure}
%
The KVA telescope is composed of a 0.6m f/15 Cassegrain devoted to
polarimetry, and a 0.35m f/11 SCT auxiliary telescope for
multicolour photometry. KVA is equipped with a polarimeter
(superachromatic half-wave retarder, with a calcite plate and a
Marconi $1K \times 1K$, 13 $\mu$m pixels, thinned back illuminated
CCD with high blue sensitivity), and a multiband photometry camera
(SBIG ST-8E, $1K \times 1.5K$ CCD, plus filter wheel with $BVRI$
filters). For other instrument details see \citet{berdyugin04}.
This telescope has been successfully operated in remote way
(mainly from Finland) through internet connection since autumn
2003. It is controlled by two PCs, incremental encoders and a
``Watchdog'' software. An IR webcam allows visual checks, a
weather station on the mountains keeps track of the observing
conditions, and an emergency shutdown procedure is active. KVA is
mainly used for optical support observations for MAGIC, long-term
monitoring of blazars, and polarimetric surveys. Different types
of observations are possible for blazars: a) one $R$-band
photometric point per available night per source; b) a
less-sampled but continuous monitoring during the year; c) $BVR$
sequences with polarization data on selected nights. A preview of
the data is available at \texttt{http://users.utu.fi/kani/1m/}.
\par
\begin{figure}[t!!]
%
\hspace{-0.5cm} \epsfysize=7.0cm %
{\epsfbox{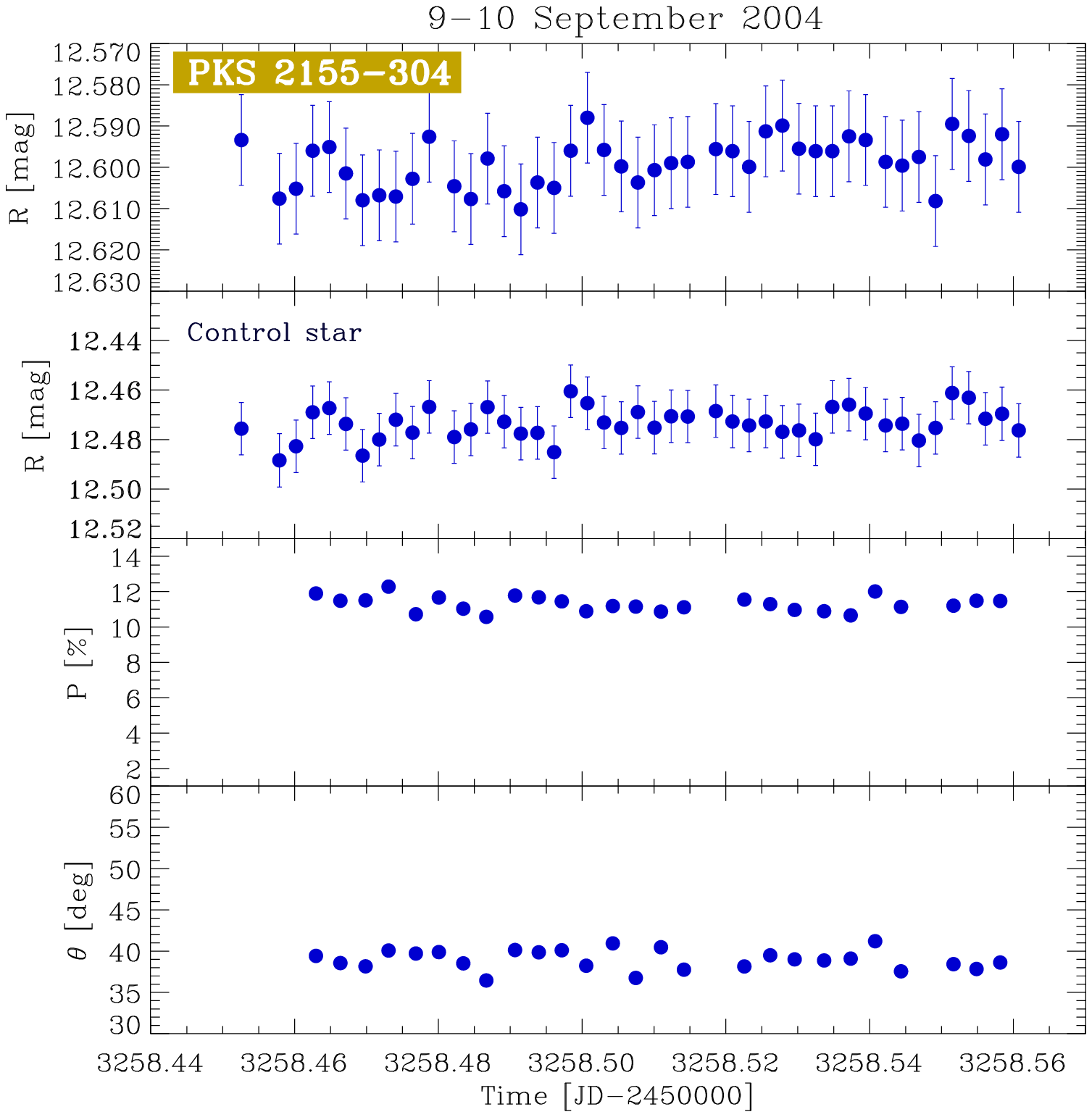}} %
\hspace{-0.2cm} \epsfysize=7.0cm %
{\epsfbox{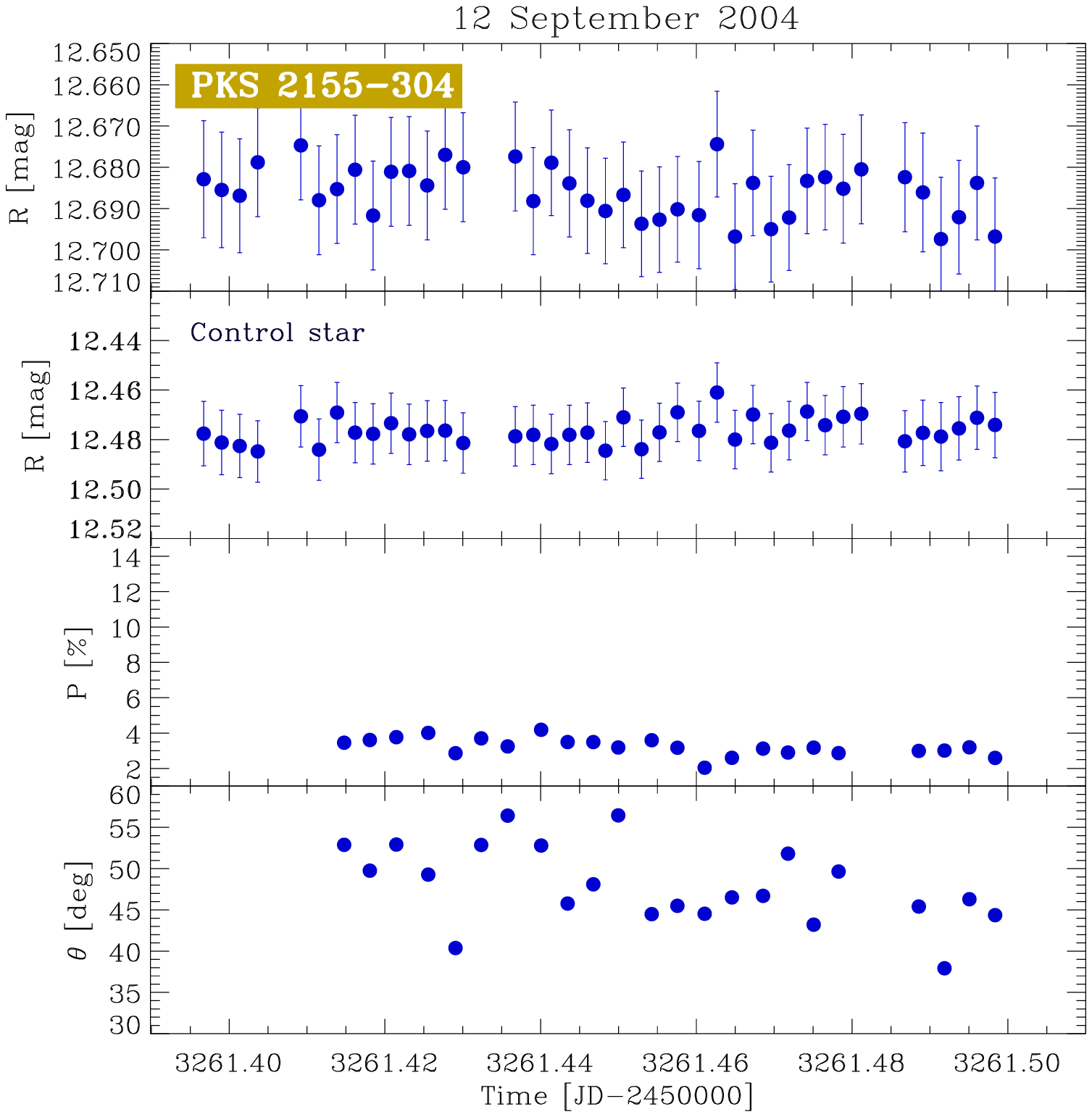}} %
\vspace{-8mm} \caption[h]{Two intranight light curves of PKS
2155-304 in $R$-magnitude, and the corresponding unfiltered
(white) linear polarization degree and position angle data. A
preliminary inspection shows no relevant intraday variability. The
decrease of the polarization degree with the flux as like the
increase of randomization/variability of the magnetic field
direction from Sep. 9-10 to Sep. 12 is clearly visible, suggesting
a decay after a non-thermal flare.}\label{figintranights}
\vspace{-2mm}
\end{figure}

 For the first time, a so low-declination blazar
like PKS 2155-304 was observed with the KVA. A total of 440
$R$-band photometric data points were obtained in 13 nights ($\sim
38$ frames per night on average) during August 3 -- September 23,
2004, in the frame of a multifrequency campaign involving HESS and
RXTE. During 4 nights (Sep. 8-9, 9-10, 12 and 22-23) simultaneous
polarimetric observations were also performed. The complete
optical light curves are presented in Fig.\ref{figlightcurves},
and two intranight light curves with polarimetric data are shown
in Fig.\ref{figintranights}.
%
%
A remarkable optical brightening (up to $R=12.548 \pm 0.012$ mag)
was observed during Sep. 8-10. A preliminary inspection on the 13
intranight light curves does not show any relevant signature of
intraday variability, but a more detailed data analysis in under
way. Since the galactic latitude of PKS 2155-304 is high ($b=
-52.246 ^{\circ} $), interstellar polarization along line of sight
is negligible. The relatively high level of optical polarization
recorded during Sep. 9-10 might provide a signature of the
synchrotron nature of this flare.
\par The KVA observations described above were also a useful preparatory
run for the optical follow-up of planned MAGIC observations of PKS
2155-304 at large zenith-angles. MAGIC should be able to derive a
proper $\gamma$-ray spectrum above 500 GeV  for it, and the
spectral cutoff due to the extragalactic background light (EBL)
absorption, with moderate pointing time and reduced systematic
effects \citep{kranich05}. PKS 2155-304 represents also an optimal
target for simultaneous observations by MAGIC and HESS. In fact
possible collaboration observations could provide a larger and
well defined energy spectrum of the source, an increased
statistics and a cross-calibration of the two detectors.
\acknowledgements 
The Tuorla Observatory group, belonging to the Research Training
Network ENIGMA, acknowledges European Community funding under
contract HPRN-CT-2002-00321. The Tuorla and KVA monitoring
programmes have been partially supported by the Academy of
Finland.
%
%
%

%

\begin{thebibliography}{}
%
\footnotesize{
%
\bibitem[Aharonian et al.(2005a)]{Aharonian05a} Aharonian, F., Akhperjanian, A. G.,
Aye, K.-M. et al.\ 2005a, A\&A, 430, 865
%
\bibitem[Aharonian et al.(2005b)]{Aharonian05b} Aharonian, F.,
Akhperjanian, A.G., Bazer-Bachi, A.R. et al. 2005b, A\&A, in press
%
\bibitem[Berdyugin et al.(2004)]{berdyugin04} Berdyugin, A.,
Piirola, V., \& Teerikorpi, P.\ 2004, A\&A, 424, 873
%
\bibitem[Brinkmann et al.(2000)]{brinkmann00} Brinkmann, W.,
Gliozzi, M., Urry, C.~M., Maraschi, L., \& Sambruna, R.\ 2000,
A\&A, 362, 105
%
\bibitem[Cagnoni et al.(2004)]{cagnoni04} Cagnoni, I., Nicastro,
F., Maraschi, L., Treves, A., \& Tavecchio, F.\ 2004, \apj, 603,
449
%
\bibitem[Chadwick et al.(1999)]{chadwick99} Chadwick, P.~M.,
Lyons, K., McComb, T. J. L., et al.\ 1999, ApJ, 513, 161
%
\bibitem[Chiappetti et al.(1999)]{chiappetti99} Chiappetti, L., Maraschi, L.,
Tavecchio, F., et al.\ 1999, ApJ, 521, 552
%
\bibitem[Ciprini(2003)]{ciprini03} Ciprini, S.\ 2003, New
Astron. Rev., 47, 709
%
\bibitem[Cortina (2005)]{cortina05} Cortina, J.\ 2005, Ap\&SS, 297,
245
%
\bibitem[Dominici et al.(2004)]{dominici04} Dominici, T.~P.,
Abraham, Z., Teixeira, R., \& Benevides-Soares, P.\ 2004, AJ, 128,
47
%
\bibitem[Edelson et al.(2001)]{edelson01} Edelson, R., Griffiths,
G., Markowitz, A., et al.\ 2001, ApJ, 554, 274
%
\bibitem[Kataoka et al.(2000)]{kataoka00} Kataoka, J., Takahashi, T.,
Makino, F., et al. \ 2000, ApJ, 528, 243
%
\bibitem[Kranich et al.(2005)]{kranich05} Kranich, D.,
Goebel, F., Laille, A., \& Wagner, R. 2005, MAGIC collab. internal
note.
%
\bibitem[Lorenz (2004)]{lorenz04} Lorenz, E.\ 2004, New Astron. Rev., 48, 339
%
\bibitem[Piner \& Edwards(2004)]{piner04} Piner, B.~G., \&
Edwards, P.~G.\ 2004, ApJ, 600, 115
%
\bibitem[Nicastro et al.(2002)]{nicastro02} Nicastro, F., Zezas, A.,
Drake, J., et al.\ 2002, ApJ, 573, 157
%
\bibitem[Tanihata et al.(2001)]{tanihata01} Tanihata, C., Urry,
C.~M., Takahashi, T. et al. 2001, ApJ, 563, 569
%
\bibitem[Tommasi et al.(2001)]{tommasi01} Tommasi, L.,
D{\'{\i}}az, R., Palazzi, E., et al.\ 2001, ApJS, 132, 73
%
\bibitem[Urry et al.(1997)]{urry97} Urry, C.~M., Treves, A., Maraschi, L.,
et al.\ 1997, ApJ, 486, 799
%
\bibitem[Vestrand \& Sreekumar(1999)]{vestrand99} Vestrand, W.~T.,
\& Sreekumar, P.\ 1999, Astropart. Phys., 11, 197
%
\bibitem[Zhang et al.(2002)]{zhang02} Zhang, Y.~H., Treves, A., Celotti,
A., et al.\ 2002, ApJ, 572, 762
%
%
%
}
%
%
\normalsize \normalsize
%
\end{thebibliography}
\end{document}